\documentclass[final]{aipproc}

\layoutstyle{6x9}

\usepackage{amsmath}

\newcommand{\kin}{k_{\text{in}}}
\newcommand{\kout}{k_{\text{out}}}

\begin{document}

\title{Improved spectral algorithm for the detection of
  network communities}

\author{Luca ~Donetti}{
  address={Departamento de Electr{\'o}nica y Tecnolog{\'\i}a de Computadores,\\
    Universidad de Granada, 18071 Granada, Spain} }

\author{Miguel ~A.~Mu\~noz}{
  address={Departamento de Electromagnetismo y F{\'\i}sica de la Materia e\\
    Instituto de F{\'\i}sica Te\'orica y Computacional \emph{Carlos I},\\
    Universidad de Granada, 18071 Granada, Spain}
}

\begin{abstract}
  We review and improve a recently introduced method for the detection
  of communities in complex networks. This method combines spectral
  properties of some matrices encoding the network topology, with well
  known hierarchical clustering techniques, and the use of the
  \emph{modularity} parameter to quantify the goodness of any possible
  community subdivision. This provides one of the best available
  methods for the detection of community structures in complex
  systems.
\end{abstract}

\maketitle

Complex networks have recently been an active topic of investigation
in physics because of their relevance in the modeling of many real
complex systems ranging from social and communication networks to
biology and neural sciences \cite{complex}. A common feature of many
of these real networks is the presence of communities, that is subsets
of nodes with high mutual interconnectivity and only few links to the
rest of the network.

The importance of their proper detection stems from many different
causes: first of all they provide a coarse-grained structure that can
notoriously simplify the analysis of a large network. Moreover,
communities can be identified as functional units in several cases of
biochemical or neural networks.  Therefore, even if there is no
commonly accepted quantitative definition of community, many
algorithms have been proposed to split a network into densely
interconnected subsets
\cite{cg,method,Duch}. A recent comparative review of most of the available 
community finding methods can be found in \cite{Danon}.  

For other problems, similar in spirit to this one, as for example
graph partitioning (in a given number of subsets), image partitioning,
or graph visualization, spectral techniques have proven to be very
useful
\cite{spectral}. Such methods are based on the spectral analysis
of a suitable matrix encoding the corresponding network topology.
 Similar techniques can also be
exploited for the detection of communities \cite{jenny,orig}.

Here we give a brief outline of a method we recently introduced
\cite{orig} which combines spectral properties, hierarchical
clustering techniques, and the optimization of the modularity (a
quantity introduced to quantify the validity of any given community
subdivision) \cite{modul}.

The nodes of a given network are represented as points in a
$D$-dimensional space whose coordinates are the components of the
first $D$ non-trivial eigenvectors of the corresponding Laplacian
matrix \cite{lapl}. Once the nodes have been embedded in a space, a
distance (Euclidean, angular, etc \cite{orig}) between them can be
defined. Afterwards, standard methods such as hierarchical clustering
techniques \cite{clustering,orig} are employed to group the nodes
according to their mutual distances: nearby sites are progressively
grouped together. Proceeding like this, a dendrogram, that is a tree
representing the hierarchy, is obtained. In order to determine at
which level the ``tree'' should be looked at to obtain the best
community-splitting, we have to quantify the quality of the
partitions.  For this purpose, the \emph{modularity} $Q$, defined as
the fraction of internal edges minus its expected value for a random
graph with the same number of links for each community, has been
introduced. The output of the algorithm is therefore the partition of
the dendrogram giving the highest value of $Q$.

The justification for using the eigenvectors of the Laplacian matrix
representing the network, can be understood by exploiting the
connection between the eigenvalue problem and the minimization of the
quadratic form
\begin{equation}
  \label{ssquares}
  \sum_{\text{links}} (x_i-x_j)^2 = x^{\text T} \mathbf{L} x ,
\end{equation}
where the $x=\{x_i\}$ is a vector of real values assigned to the nodes
and $\mathbf{L}$ is the Laplacian matrix \cite{lapl}.  Minimizing this
expression is a way to impose the condition that connected nodes
should be given a similar value of $x$. Indeed, it is easy to see that
minimizing equation~\eqref{ssquares} with a normalization condition on
vector $x$ ($\sum x_i^2 = 1$) yields the eigenvalue equation for
matrix $\mathbf{L}$. The first eigenvector is trivial (constant) and
the corresponding eigenvalue is zero: actually if all $x_i$ are equal
the sum \eqref{ssquares} is zero and it is its minimum possible
value. The following eigenvector (with an eigenvalue larger than $0$
for any connected network) corresponds to the non-trivial minimum and
therefore its components can be used to partition the nodes. Indeed,
as shown in \cite{orig}, also the following eigenvectors contain
useful information and can be profitably used to find communities in
the network. The number of eigenvectors $D$ that have to be taken
into account in order to obtain a good detection of communities is
\emph{a priori} not known.
 Therefore, the whole
procedure is repeated in the algorithm for different $D$'s and the
subdivision corresponding to the highest value of the modularity is
selected.

If we assign to the nodes a weight proportional to their degree, the
normalization condition becomes $\sum k_i x_i^2 = x^{\text T}
\mathbf{D} x = 1$; in this case the minimization of
equation~\eqref{ssquares} is transformed into the eigenvalue equation
for the matrix $\mathbf{L'} = \mathbf{D^{-1}L}$. As before, the first
non trivial eigenvector corresponds to the non-trivial minimum of the
sum \eqref{ssquares}. Therefore, we can wonder how the original method
performance (as presented in \cite{orig}) is affected by replacing the
eigenvectors of $\mathbf{L}$ by those of $\mathbf{L'}$.

First of all, we applied both algorithms (with $\mathbf{L}$ and with
$\mathbf{L'}$  respectively) 
\footnote{An implementation of the algorithms can be
  found at \texttt{http://www.ugr.es/$\,\tilde{\;}$donetti/}.} to
  computer generated networks with a given community structure
\cite{cg}. These networks contain $128$ nodes, split into $4$
equal-size communities; edges are randomly extracted in such a way
that each node has, on the average, $\kin$ links to other nodes in the
same community and $\kout$ to to the rest of the network, with $\kin +
\kout = 16$.  For small $\kout$ the communities are almost
disconnected, while increasing this value they become less and less
separated, so that detecting them becomes a very difficult and not
clearcut task. Since the communities are known, we can measure the
quality of the algorithm by counting the number of nodes that are
correctly classified.
\begin{figure}[tb]
  \centering \includegraphics[width=.8\textwidth]{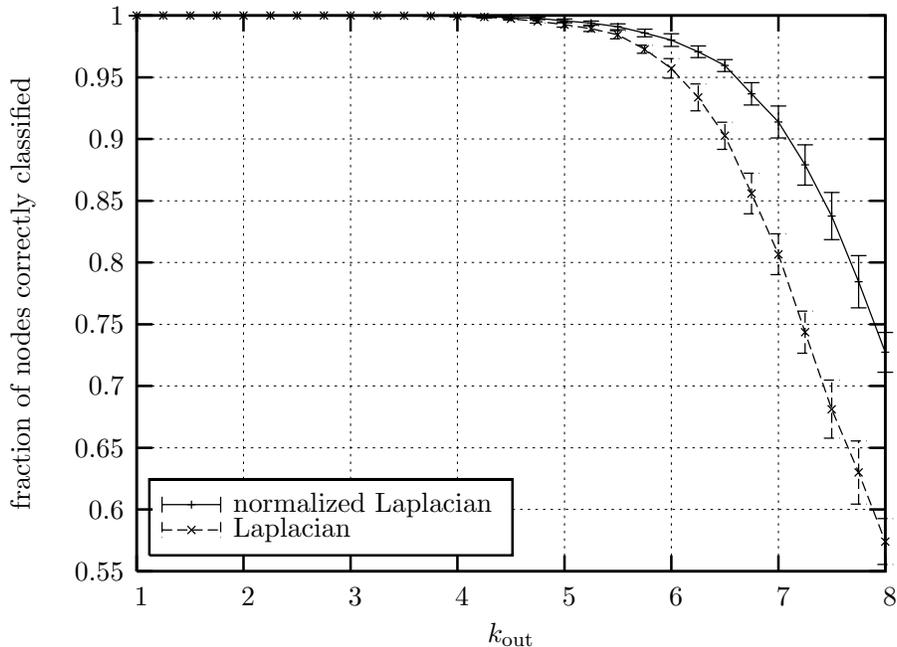}
  \caption{Fraction of nodes correctly classified by the algorithm
  (averaged over 200 networks) as a function of $\kout$, using the
  eigenvectors of $\mathbf{L}$ and $\mathbf{L}'$. In both cases
  angular distance and complete linkage clustering are used (see
  \cite{orig}).  } \label{fraction}
\end{figure}
In figure~\ref{fraction} we plot the corresponding fraction of nodes,
and we can see that when the eigenvectors of the normalized Laplacian
matrix $\mathbf{L'}$ are used, the method produces much better
results.  Moreover, in a very recent independent and systematic
comparison of different community-finding methods performed by Danon
et al. \cite{Danon}, it has been found that our method, equipped with
the normalized Laplacian matrix, exhibits an extremely good
performance and is among the most convenient choices.

Another network which is used as a test for many community finding
algorithm is the Zachary karate club \cite{zachary}. In this case, we
can compare the modularity value corresponding to the best split in
the two cases: using the Laplacian eigenvectors we obtain $Q=0.412$
while using the eigenvectors of $\mathbf{L'}$ leads to $Q=0.419$ which
is the best value obtained so far for such a workbench problem
\cite{Duch}.

 As a last example, we have studied the jazz bands network
 \cite{jazz}, which is also one of the prototypical instances studied
 in this field. Using  the Laplacian we measure $Q= 0.437$, while with
$\mathbf{L'}$ the modularity increases to $Q=0.444$ (almost identical 
to the best available result \cite{Duch}).

Summarizing, we outlined the connection between the detection of
communities and the spectral properties of some proper matrices
describing the network topology. Moreover, we improved the performance
of the algorithm described in \cite{orig} by using the eigenvectors of
a different matrix: the normalized Laplacian matrix. We do not have a
clear understanding of why the method equipped with this matrix gives
better results than with the Laplacian matrix, but as a matter of fact
this is actually the case in all the tested examples. Finally, let us
mention that the method (with either matrix) can be easily generalized
to the case of weighted networks.

\begin{theacknowledgments}
We are thankful to L. Danon and A. Arenas for a useful exchange of
correspondence, and to M. Verbeni for a critical reading of the
manuscript. Also, financial support from the Spanish MCyT (FEDER)
under project BFM2001-2841 and the EU COSIN project IST2001-33555 is
acknowledged.
\end{theacknowledgments}

\end{document}